\documentclass[]{spie}
\title{Optical Solitons in an Anisotropic Medium with Arbitrary Dipole Moments}
\author{N. V. Ustinov
\skiplinehalf
Quantum Field Theory Department, Tomsk State University,\\
36 Lenin Avenue, Tomsk, 634050, Russia}
\authorinfo{Further author information: 
E-mail: n\_ustinov@mail.ru}
\begin{document}
\maketitle
\begin{abstract}
We find the Lax pair for a system of reduced Maxwell--Bloch equations that 
describes the propagation of two-component extremely short electromagnetic 
pulses through the medium containing two-level quantum particles with 
arbitrary dipole moments. 
\end{abstract}

\noindent

\section{Introduction}
Generation of extremely short pulses\cite{FB_CBSh,DHZA,SSKS,BWPW,NDe_SSSFSSK} 
(ESP) with a duration of a few periods of light oscillations has offered a 
strong incentive for theoretical studies of their interaction with matter (see 
reviews\cite{BK,M} and references therein). 
For obvious reasons, the slowly varying envelope approximation commonly 
exploited in the nonlinear optics of quasi-monochromatic (or ultrashort) pulses 
cannot be applied to investigate the propagation of ESP. 

The slowly varying envelope approximation was not used in the case of the 
ultrashort pulses in Ref.~\citenum{EGCB}, where an alternative approach to the 
theory of self-induced transparency\cite{McCH,Lamb1} (SIT) was developed. 
This approach was based on the assumption of low density of the quantum 
particles, which was consistent with conditions of the SIT experiments. 
Since the backscattered wave is weak in that case, the order of derivatives in 
the wave equation for the pulse field can be reduced by using the 
unidirectional propagation (UP) approximation\cite{ElB}. 
The resulting so-called reduced Maxwell-Bloch (RMB) equations\cite{EGCB}\,, as 
well as the SIT equations\cite{Lamb1}\,, are integrable by the inverse 
scattering transformation (IST) method\cite{Lamb2,BC,ZMNP}. 
This method is widely recognized as one of the most powerful tools in studying 
the nonlinear phenomena. 
In particular, the pulse dynamics in the integrable models of nonlinear optics 
is described by the soliton solutions. 

In the last years, the theoretical investigation of coherent nonlinear effects 
in anisotropic media attracts great 
attention\cite{Kas,AEGM,S1,MC,Elyutin,SU1,SU2,SZM,U,BU,SU3,SU4}. 
This is caused by the significant development of the nanotechnologies and the 
methods of producing the low-dimensional quantum structures. 
Unlike the case of isotropic media, the parity of the stationary states of the 
quantum particles of the anisotropic medium is not well defined. 
That is why the diagonal elements of the matrix of the dipole moment operator
and their difference, which is called the permanent dipole moment (PDM) of the
transition, may not vanish. 
An optical pulse propagating in such a medium not only induces transitions 
between these states, but also shifts the transition frequency via dynamic 
Stark effect. 
Owing to this, the ultrashort pulses can propagate in the medium in the 
regimes that differ from the SIT one\cite{S1,SU1}. 

The dynamics of one-component ESP in the anisotropic media was studied in 
papers\cite{Kas,AEGM,MC,Elyutin,SU2,U}\,. 
It was revealed\cite{AEGM} that the scalar RMB equations with PDM are 
integrable in the frameworks of the IST.
The numerical investigation of the pulse formation governed by these equations 
displayed an existence of solitary stable bipolar signal with nonzero time 
area\cite{Elyutin}. 
Its solitonic nature was established in Refs.~\citenum{SU2} and \citenum{U}. 

The propagation through anisotropic media of the two-component ESP was 
investigated in\cite{SZM,BU,SU3,SU4}\,. 
The case, where one of the pulse components causes the quantum transitions, 
while another one shifts the energy levels, was considered in 
Ref.~\citenum{BU}. 
Corresponding two-component RMB equations differ by notations from the system 
describing the transverse-longitudinal acoustic pulse propagation in the 
low-temperature paramagnetic crystal and are integrable\cite{Z1}. 
An application of the spectral overlap approximation to these equations gives 
one more integrable model\cite{SU3,SU4}. 
If both components of the ESP excite the quantum transitions only (i.e., PDM 
of the transition is equal to zero), the proper system of the two-component 
RMB equations is also integrable\cite{SZM}. 

We see that different particular cases of the two-component RMB equations are 
integrable by means of the IST method. 
The main aims of the present study are to join these cases and to find more 
general conditions, under which the integrability of the RMB equations takes 
place. 

\section{Two-component system of the Maxwell--Bloch equations}

Let us consider the medium containing two-level quantum particles.
Assume for simplicity that the medium is isotropic, i.e. an anisotropy is 
induced by the quantum particles. 
Let the plane ESP propagate through the medium in the positive direction of 
$y$ axes of the Cartesian coordinate system. 
Then the Maxwell equations yield the following system for projections $E_x$ 
and $E_z$ of the electric field: 
\begin{equation}
\frac{\partial^2E_x}{\partial y^2}-\frac{n^2}{c^2}\frac{\partial^2E_x}
{\partial t^2}=\frac{4\pi}{c^2}\frac{\partial^2P_x}{\partial t^2}, 
\label{E_x_yy}
\end{equation}
\begin{equation}
\frac{\partial^2E_z}{\partial y^2}-\frac{n^2}{c^2}\frac{\partial^2E_z}
{\partial t^2}=\frac{4\pi}{c^2}\frac{\partial^2P_z}{\partial t^2},
\label{E_z_yy}
\end{equation}
where $P_x$ and $P_z$ are the components of polarization connected with the 
two-level particles; $n$ is the refractive index of the medium; $c$ is the 
speed of light in free space. 

To describe the evolution of quantum particles, we exploit the von~Neumann 
equation for density matrix $\hat\rho$:
\begin{equation}
i\hbar\frac{\partial\hat\rho}{\partial t}=[\hat H,\hat\rho].
\label{rho_t}
\end{equation}
Here Hamiltonian $H$ of the two-level particle is written as follows 
\begin{equation}
\hat H=\mbox{diag}(0,\hbar\omega_0)-\hat d_xE_x-\hat d_zE_z, 
\label{Ham}
\end{equation}
where $\omega_0$ is the resonant frequency of quantum transition; $\hat d_x$ 
and $\hat d_z$ are the matrices of the projection of the dipole moment 
operator on $x$ and $z$ axes, respectively; $\hbar$ is the Plank's constant. 

The expressions for the polarization components read as 
\begin{equation}
P_x=N\,\mbox{Tr}\,(\hat\rho\hat d_x),
\label{P_x}
\end{equation}
\begin{equation}
P_z=N\,\mbox{Tr}\,(\hat\rho\hat d_z),
\label{P_z}
\end{equation}
where $N$ is the concentration of the quantum particles.
We suppose that the matrices of the dipole moment are defined as given 
\begin{equation}
\hat d_x=\left(
\begin{array}{cc}
D_x&d_1\\
d_1&0
\end{array}
\right)\!,
\label{d_x}
\end{equation}
\begin{equation}
\hat d_z=\left(
\begin{array}{cc}
D_z&\delta+id_2\\
\delta-id_2&0
\end{array}
\right)\!,
\label{d_z}
\end{equation}
where $d_1$, $d_2$, $\delta$, $D_x$ and $D_z$ are real parameters. 
This representation for the dipole moment matrices is general, but we have 
reduced it to a simpler form. 
Quantities $D_x$ and $D_z$ are nothing but the PDM projections. 

Using equations (\ref{rho_t})--(\ref{d_z}), we find 
\begin{equation}
\frac{\partial W}{\partial t}=2\frac{d_2}{\hbar}\,E_zU-2\left(
\frac{d_1}{\hbar}\,E_x+\frac{\delta}{\hbar}\,E_z\right)V,
\label{W_t}
\end{equation}
\begin{equation}
\frac{\partial U}{\partial t}=-\left(\omega_0+\frac{D_x}{\hbar}\,E_x+
\frac{D_z}{\hbar}\,E_z\right)V-2\frac{d_2}{\hbar}\,E_zW, 
\label{U_t}
\end{equation}
\begin{equation}
\frac{\partial V}{\partial t}=\left(\omega_0+\frac{D_x}{\hbar}\,E_x+
\frac{D_z}{\hbar}\,E_z\right)U+2\left(\frac{d_1}{\hbar}\,E_x+
\frac{\delta}{\hbar}\,E_z\right)W, 
\label{V_t}
\end{equation}
where 
$$
W=\frac{\rho_{22}-\rho_{11}}{2},\quad
U=\frac{\rho_{12}+\rho_{21}}{2},\quad
V=\frac{\rho_{12}-\rho_{21}}{2i} 
$$
are the components of the Bloch vector; $\rho_{jk}$ ($j,k=1,2$) are the 
elements of the density matrix. 

Let the concentration of the quantum particles be small: 
$N\|\hat d_{x,\,z}\|^2/\hbar\omega_0\!\ll\!1$, where $\|C\|$ is the norm of 
the matrix $C$. 
Then we are able to reduce the order of derivatives in the wave equations for 
the electric field components. 
An application of the UP approximation\cite{ElB} to equations (\ref{E_x_yy}), 
(\ref{E_z_yy}) and exclusion of the time derivatives of the elements of the 
density matrix with the help of (\ref{W_t})--(\ref{V_t}) give 
\begin{equation}
\frac{\partial E_x}{\partial y}+\frac{n}{c}\frac{\partial E_x}{\partial t}=
\frac{4\pi N}{nc\hbar}\Bigl[SE_z+\hbar\omega_0d_1V\Bigr],
\label{E_x_y}
\end{equation}
\begin{equation}
\frac{\partial E_z}{\partial y}+\frac{n}{c}\frac{\partial E_z}{\partial t}=-
\frac{4\pi N}{nc\hbar}\Bigl[SE_x+\hbar\omega_0(d_2U-\delta V)\Bigr], 
\label{E_z_y}
\end{equation}
where $S=2d_1d_2W+d_2D_xU+(d_1D_z-\delta D_x)V$. 

The two-component system of RMB equations (\ref{W_t})--(\ref{E_z_y}) 
describes the propagation of vector ESP in the medium containing two-level 
quantum particles with arbitrary dipole moments. 
It is seen that both electric field components, as well as any superposition 
of them, fulfill two different functions in the general case: they excite the 
quantum transitions and shift the energy levels due to PDM. 
Obviously, the system obtained coincides with the RMB equations for isotropic 
medium\cite{EGCB} if we put PDM equal to zero ($D_x=D_z=0$) and $E_z=d_2=0$ 
(or $E_x=d_1=0$). 
Other integrable cases of equations (\ref{W_t})--(\ref{E_z_y}) were studied 
in\cite{AEGM,SZM,BU}\,. 

\section{Lax pair}

An integrability of the nonlinear equations given by means of the IST method 
implies an opportunity to represent them as the compatibility condition of 
the overdetermined system of linear equations (Lax pair). 
Consider the following Lax pair 
\begin{equation}
\begin{array}{l}
\displaystyle\frac{\partial\psi}{\partial t_{\mathstrut}}=
L(\lambda)\psi(\lambda),\\ 
\displaystyle\frac{\partial\psi^{\mathstrut}}{\partial y}=
A(\lambda)\psi(\lambda),
\end{array}
\label{Lax}
\end{equation}
where $\lambda$ is so-called spectral parameter; $L(\lambda)$ and $A(\lambda)$ 
are matrices; $\psi=\psi(y,t,\lambda)$ is a solution of the overdetermined 
system. 
Its compatibility condition reads as 
\begin{equation}
\frac{\partial L(\lambda)}{\partial y}-\frac{\partial A(\lambda)}{\partial t}
+[L(\lambda),A(\lambda)]=0. 
\label{cc}
\end{equation}

Using the overdetermined linear systems for the cases discussed 
in\cite{AEGM,SZM}\,, we can offer possible form of the Lax pair for 
(particular case of) the equations (\ref{W_t})--(\ref{E_z_y}). 
This form contains coefficients in matrices $L(\lambda)$ and $A(\lambda)$, 
which are a subject of subsequent definition, and may be valid only under 
imposing some constraints on the elements of matrices (\ref{d_x}) and 
(\ref{d_z}) of the dipole moment projections. 
Having written down the compatibility condition and having excluded 
the derivatives with the help of (\ref{W_t})--(\ref{E_z_y}), we obtain the 
overdetermined system of algebraic equations on the entered coefficients. 
The number of the coefficients should be great enough to include into a 
consideration both the cases we start with. 
For this reason, the system of the algebraic equations is strongly 
overdetermined. 
Nevertheless, we have been able to solve this system after straightforward, 
but tedious calculations. 
Moreover, it has been done without imposing additional constraints on the 
dipole moments of the transition. 
We have found the following expressions for matrices $L(\lambda)$ and 
$A(\lambda)$ of system (\ref{Lax}): 
\begin{equation}
L(\lambda)=\left( 
\begin{array}{cc}
\displaystyle\frac{i}{2}\left[\lambda^2-\frac{b}{\lambda^2}\right]&
\displaystyle\frac{1}{2\sqrt2\hbar}\left[\lambda E^*+\frac{\delta_2}{\delta_1}
\frac{E}{\lambda}\right]_{\mathstrut}\\
\displaystyle\frac{\delta_1}{2\sqrt2\hbar}\left[\lambda E+
\frac{\delta_2^*}{\delta_1}\frac{E^*}{\lambda}\right]^{\mathstrut}&
\displaystyle-\frac{i}{2}\left[\lambda^2-\frac{b}{\lambda^2}\right]
\end{array}
\right),
\label{L}
\end{equation}
\begin{equation}
A(\lambda)=\frac{2\pi N}{nc}\frac{1}{\displaystyle\lambda^2+\frac{b}
{\lambda^2}+B}
\left(
\begin{array}{cc}
\displaystyle\frac{i}{\hbar}\left[\lambda^2-\frac{b}{\lambda^2}\right]S&
\displaystyle\frac{\sqrt2d_1d_2}{\hbar^2\delta_1}\left[\lambda Q^*+
\frac{\delta_2}{\delta_1}\frac{Q}{\lambda}\right]_{\mathstrut}\\     
\displaystyle\frac{\sqrt2d_1d_2}{\hbar^2}\left[\lambda Q+
\frac{\delta_2^*}{\delta_1}\frac{Q^*}{\lambda}\right]_{\mathstrut}&
\displaystyle-\frac{i}{\hbar}\left[\lambda^2-\frac{b}{\lambda^2}\right]S 
\end{array}
\right)-\frac{n}{c}\,L(\lambda), 
\label{A}
\end{equation}
where 
$$
E=E_x+iE_z+\frac{\delta_3}{\delta_1},\quad Q=\delta_3W+\delta_4U+\delta_5V,
$$
$$
\delta_1=\frac{2}{\omega_0d_1d_2}\Bigl[(4d_1^2+D_x^2)d_2^2+
(d_1D_z-\delta D_x)^2\Bigr], 
$$
$$
\delta_2=2d_2^2+2(\delta+id_1)^2+\frac{(D_z+iD_x)^2}{2},
$$
$$
\delta_3=\frac{2\hbar}{d_1d_2}\Bigl[d_2^2D_x+
(\delta-id_1)(\delta D_x-d_1D_z)\Bigr], 
$$
$$
\delta_4=\frac{\hbar}{d_1d_2}\Bigl[\delta D_xD_z-d_1(4d_2^2+D_z^2)+
i(d_1D_z-\delta D_x)D_x\Bigr], 
$$
$$
\delta_5=\frac{\hbar}{d_1}\Bigl[4d_1\delta+D_xD_z-i(4d_1^2+D_x^2)\Bigr], 
$$
$$
b=\frac{|\delta_2|^2}{\delta_1^2},\quad
B=\frac{1}{\delta_1}\Bigl[4(d_1^2+d_2^2+\delta^2)+D_x^2+D_z^2\Bigr]. 
$$
Substituting (\ref{L}) and (\ref{A}) into (\ref{cc}), we see that the equality 
is fulfilled only if equations (\ref{W_t})--(\ref{E_z_y}) take place. 
Thus, the two-component RMB equations (\ref{W_t})--(\ref{E_z_y}) belong to the 
class of nonlinear models integrable by means of the IST method at {\bf any} 
values of parameters $d_1$, $d_2$, $\delta$, $D_x$ and $D_z$. 

Equations (\ref{cc}), (\ref{L}) and (\ref{A}) with $\delta=D_x=D_z=0$ give the 
Lax pair presented in Ref.~\citenum{SZM}. 
A connection of the Lax pair we found with the pair obtained in 
Ref.~\citenum{AEGM} is not so obvious since $d_1d_2=0$ in the last case. 

\section{Conclusion}

We have established the integrability in the frameworks of the IST method of 
the system of two-component RMB equations for anisotropic medium in the most 
general case. 
This implies that these equations have multi-soliton solutions, Darboux and 
B\"acklund transformations, infinite hierarchies of the conservation laws and 
infinitesimal symmetries and other attributes of the integrable models. 
The detailed study of them is a subject of further researches. 

\section*{ACKNOWLEDGMENT}

This work was supported by the Russian Foundation for Basic Research (grant 
No.\,\,05--02--16422).

\end{document}